# Laser excitation of gigahertz vibrations in Cauliflower mosaic viruses' suspension

N. V. Tcherniega[1], S. M. Pershin[2], A. F. Bunkin[2], E. K. Donchenko[3], O. V. Karpova[3], A. D. Kudryavtseva[1], V. N. Lednev[2], T. V. Mironova[1], M. A. Shevchenko[1], M. A. Strokov[1], and K. I. Zemskov[1]

[1]*P.N. Lebedev Physical Institute of the RAS, Moscow, Russia*
[2]*A.M. Prokhorov General Physics Institute of the RAS, Moscow, Russia*
[3]*M.V. Lomonosov Moscow State University, Moscow, Russia*

**Abstract:** The interaction of laser pulses with the Cauliflower mosaic virus (CaMV) in a Tris-HCl pH7.5 buffer is investigated. 20 ns ruby laser pulses are used for excitation. Spectra of the light passing through the sample and reflected from it are registered with the help of a Fabri-Perot interferometer. Stimulated low-frequency Raman scattering (SLFRS) in a CaMV suspension is registered. The SLFRS frequency shift, conversion efficiency and threshold are measured for the first time, to the best of our knowledge.

## 1. Introduction

The excitation of vibrational states in monodispersive biological systems consisting of nanosized particles - for example, viruses - can have a significant impact on these systems up to their mechanical destruction [1]. The most efficient excitation of oscillations in such systems can be implemented with the resonant impact on the system; namely the coincidence of the system eigenfrequency with a frequency of external influence. Acoustic eigenfrequencies of nano and submicron biological objects are in the gigahertz or terahertz range and defined by shape, size and elastic properties of the systems under consideration. Resonant microwave absorption [2], acoustic radiation in giga or terahertz range [3] and impulsive Stimulated Raman Scattering (ISRS) [4-7] – are among the suggestions that are regarded as having the most possible resonant impacts on biological systems.

The strong absorption of electromagnetic radiation in the microwave range in water and a small propagation length of acoustic waves of gigahertz range make it very difficult to use the first two methods for influencing real biological systems. The use of electromagnetic radiation in the visible or near-infrared range for the efficient excitation of the vibrational modes of biological objects allows for the avoidance of the strong absorption of aqua environment. It also allows for an increase in the radiation penetration into the sample. The biharmonic radiation of nanosecond duration [8] as well as ISRS under femtosecond excitation [4-7] can realize the coherent excitation of low frequency Raman-active modes of different nanoparticles. The basic condition for the effective usage of biharmonic pumping for the efficient excitation of nanoparticle vibrations is the equality of the difference frequency and the eigenfrequency. For an exact eigenfrequency value calculation, a liquid drop model [9] or an elastic sphere model [10,11] can be used. Both approaches require some knowledge of the elastic characteristics of the nanoparticles and the environment. For the experimental eigenfrequency measurements, low frequency Raman scattering (LFRS) [12-14] can be used. Today there are a relatively small number of works [15,16], in which the low-frequency Raman scattering by viruses in the liquid environment has been obtained experimentally. It is connected with virus vibration damping due to the radiation of the acoustic energy in environment [17]. One other way to obtain information about the frequency of the acoustic vibrations of a nanoparticles system, including viruses, may be the process of stimulated low-frequency Raman scattering (SLFRS) [18-20]. SLFRS is a stimulated analogue of low-frequency Raman scattering which was realized for large number of nanoparticles of different natures. The frequency shift of SLFRS is defined by the nanoparticles dimensions and also their - and environment`s - elastic properties. Also SLFRS can be used as a highly stable, powerful source of biharmonic radiation with the difference frequency tuning range from a few gigahertz to some terahertz. The purpose of this research is to realize the SLFRS excitation in Cauliflower mosaic virus (CaMV) suspensions and to define the threshold value and conversion efficiency.

## 2. Experimental

### 2.1 Samples

CaMV is a type member of the genus Caulimovirus, of the family of Caulimoviridae. Icosahedral CaMV virions are built from 420 capsid protein subunits. Viral particles have a diameter of about 35 nm and contain a single molecule of circular double-stranded DNA. The DNA appears to replicate in the nuclei of infected cells as a plasmid. For further analysis a virus was placed in a Tris-HCl buffer, pH 7.5. The number of particles in the sample was analyzed by nanoparticle tracking analysis (NTA) according to [21,22]. The CaMV concentration was about $10^{14}$ particles/cm$^3$. A CaMV image was obtained with the help of transmission electron microscopy (figure 1).

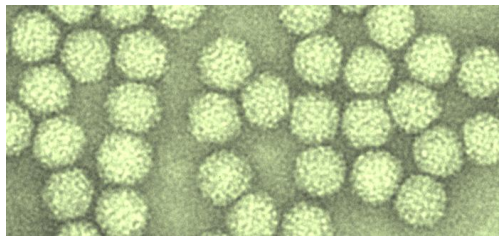

Figure 1. Transmission electron microscopy image of CaMV in Tris-HCl buffer, pH 7.5.

### 2.2 Experimental setup

For SLFRS excitation ruby laser ($\lambda$ = 694.3 nm, $\tau$ = 20 ns, $E_{max}$ = 0.3 J, $\Delta\nu$ = 0.015 cm$^{-1}$, divergence 3.5·10$^{-4}$ rad) was used. Laser light was focused at the center of the 1 cm quartz cell with a sample by the lens with a focal length of 5 cm. SLFRS spectra have been registered with Fabri-Perot interferometers with a range of dispersion 2.5 cm$^{-1}$ (75 GHz) and with a spectral resolution of about 0.06 cm$^{-1}$. The spectral registration was realized simultaneously for light scattered in forward and backward directions. An additional mirror was used in the measurement of the spectrum of the scattered radiation in the reverse direction (backward SLFRS). For comparison, the same quartz cell filled only with a Tris-HCl pH7.5 buffer was used as a reference sample. Simultaneously, the spectral measurements, conversion efficiency and the SLFRS threshold were measured. For this purpose, the total light energy transmitted through and reflected from the cell was measured by calibrated photodiodes. For the relative SLFRS intensity definition the blackening marks were used. Experimental setup is presented in the figure 2.

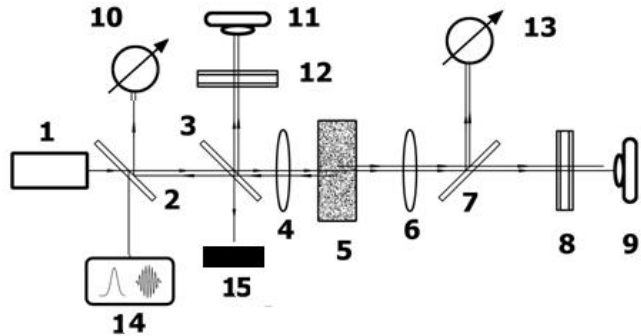

Figure 2. Experimental setup: 1 – ruby laser; 2, 3, 7 – glass plates; 5 – quartz cell with the sample; 4, 6 – lenses; 8, 12 – Fabri-Perot interferometers; 9, 11 - photo cameras, registering SLFRS spectra; 10, 13 – systems for SLFRS energy measuring in forward and backward direction; 14 – system for laser pulse characteristics measurements; 15 – mirror.

*2.3 Experimental results*

No scattering propagating in a forward direction was registered in the pure Tris-HCl pH7.5 buffer. Stimulated Brillouin scattering (SBS) propagating in a backward direction was registered in the Tris-HCl pH7.5 buffer. While the laser intensity exceeded the threshold value, SLFRS was excited in CaMV in the Tris-HCl pH7.5 buffer. The SLFRS frequency shift was measured to be 1.94 cm$^{-1}$. Figure 3 demonstrates the absence of SLFRS at the laser intensity below the threshold (figure 3 a) and the SLFRS appearance, while the laser intensity exceeds the threshold value (figure 3 b).

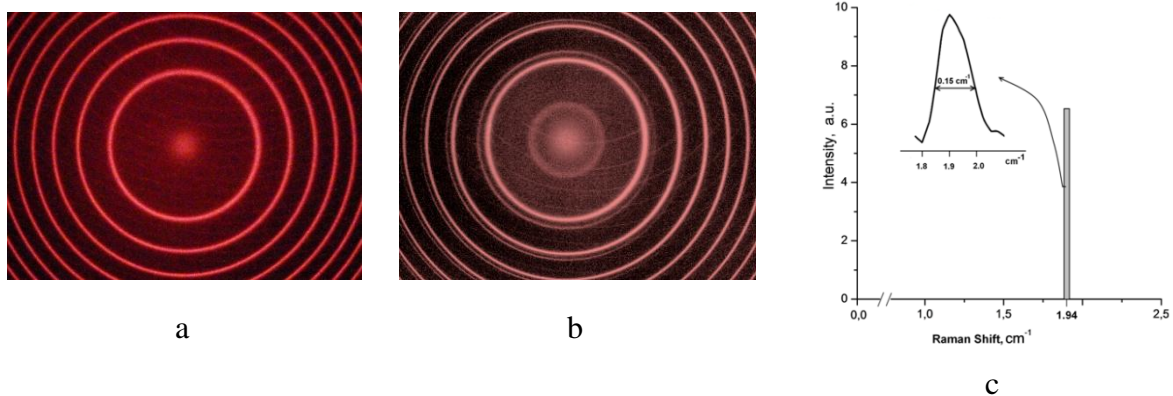

Figure 3. Fabry–Perot (range of dispersion 2.5 cm$^{-1}$ (75 GHz))- interferograms corresponding to the forward scattered radiation in the CaMV suspension in Tris-HCl pH7.5 buffer for laser intensities (a) 0.05 GW cm$^{-2}$ (below the threshold), (b) 0.12 GW cm$^{-2}$ (above the threshold) and (c) result of the interferogram digital image processing for SLFRS line.

The SLFRS threshold for scattering propagating in backward and forward directions was about 0.1 GW cm$^{-2}$. The maximum conversion efficiency of the pumping light into the scattered light was 20 per cent. The maximum intensity value, which can be stored in this acoustic excitation, is defined by the ratio of the acoustic and laser frequencies and is of the order of a 10$^{-4}$ laser intensity.

We note that the frequency shifts of the scattered waves propagating in backward and forward direction were the same (1.94 cm$^{-1}$). This fact is direct evidence that this scattering is Raman – type process, which is defined by the interaction of the laser pulse with localized nanosized virus oscillations, in contrast to the process of stimulated Brillouin scattering (SBS), which can be excited only in a backward direction (towards the laser). In our experimental conditions SBS in CaMV suspensions in a Tris-HCl pH7.5 was not excited.

## 3. Discussion

The calculation of the SLFRS shift - which corresponds to the frequency of the virus nanoparticle oscillation as for spontaneous LFRS - can be realized by using two approaches [9]: the first is based on a liquid drop model and the second is based on an elastic sphere model. In any case, for an exact frequency definition, the elastic properties of the virus under consideration must be well known, but in practice these parameters are not accurately determined. Considering the virus as a free spherical particle, one can determine the sound speed. In the case of a free nanosphere, the mode frequency is inversely proportional to the particle size D through the simplified relation $v \approx D^{-1} v_l$, where $v_l$ - speed of sound. In our experimental conditions for D = 35 nm and $v$ = 1.94 cm$^{-1}$ the sound speed in CaMV is about 2037 ms$^{-1}$.

The SLFRS conversion efficiency and threshold are defined by the morphology of the nanoparticles, and by the elastic properties of the environment namely the Tris-HCl pH7.5. For an exact definition of the sound speed in Tris-HCl pH7.5 buffer, we used the measurement of the spectral shift of the SBS first Stokes component. Using the experimental setup shown in figure 2, we determined the spectral shift of the SBS excited in the Tris-HCl pH7.5. SBS was excited only in a backward direction as it must be. SBS frequency shifts were estimated to be 0.22 cm$^{-1}$. According to equation $v = \Omega c / 2n\omega$, where $\Omega$ is frequency shift of the scattered wave, ω is the laser frequency, c is the light speed in the vacuum, n is the refraction index, the sound speed in Tris-HCl pH7.5 is 1637 ms$^{-1}$. The different elastic properties of the liquids lead to different acoustic impedance at the border of the nanoparticle and can lead to different excitation conditions of SLFRS and its conversion efficiency.

The buildup of SLFRS can be considered by analogy with Raman scattering by molecular vibrations. Let a monochromatic electromagnetic wave fall on the dielectric nanosphere. The nanoparticle makes thermal oscillations with a frequency determined by its morphology. We consider that the nanoparticle size is small in comparison with the wavelength, so that the electric field near the nanoparticle is almost uniform. The nanoparticle is polarized and acquires a certain dipole moment, oscillating in time with the eigenfrequency of the nanoparticle. The electric field produced by a dipole-polarized nanoparticle, when interacting with the field of the initial radiation, leads to low – frequency Raman scattering, in a manner similar to that for intramolecular vibrations. The frequency shift of the oscillating dipole moment is defined by the shape, size and elastic nanoparticle properties. The spontaneously inelastically scattered light that arises from nanoparticles vibrating independently is the source for the SLFRS initiation. The amplification of spontaneously scattered Stokes photons during propagation through the active medium leads to the formation of coherent directed intense Stokes radiation in the directions towards and along the pumping. SLFRS, which is formed in such systems, is a source of bi-harmonic electromagnetic radiation with a difference frequency exactly corresponding to the eigen acoustic frequency of the nanoparticles making up this system. Next, this two-frequency radiation can be significantly amplified, for example in a laser cavity, and can be used for resonant selective action on systems that have their own acoustic frequencies coinciding with the difference frequency. The physical mechanism that provides an effective impact for biharmonic pumping on the systems under study is a ponderomotive interaction. As well known, a dielectric nanoparticle in the electric field gets dipole moment which can be written in the form:

$$\vec{P} = n_1^2 \left( \frac{n^2 - 1}{n^2 + 2} \right) R^3 \vec{E} \quad (1)$$

where $n_1$ is refraction index of the surrounding medium, $n = n_2/n_1$, $n_2$– refraction index of the nanoparticle, $R$ – nanoparticle radius.

The ponderomotive force $\vec{f}$ acting from the electromagnetic field on the nanoparticle has the form:

$$\vec{f} = (\vec{P} \cdot \nabla) \vec{E} \quad (2)$$

If the system of nanoparticles is placed into a given field $\vec{E}$ consisting of two waves

$$\vec{E} = \frac{1}{2} \vec{E}_0 e^{i\omega t} + \frac{1}{2} \vec{E}_S e^{i(\omega - \Omega)t} + c.c. \quad (3)$$

the ponderomotive force will have a component oscillating with frequency $\Omega$.

$$f \approx E_0 E_S^* e^{i\Omega t} \quad (4)$$

The ponderomotive force (4) will excite harmonic acoustic vibrations in the nanoparticle. Resonant buildup of surface oscillations of a liquid drop of submicron size by an electromagnetic field was theoretically considered in [23]. It was shown that the effective impact on a liquid submicron particle can be realized by biharmonic laser radiation if the difference frequency proves to be equal to the particle's acoustic eigenfrequency. The amplitude of oscillation is determined by the following expression [23]

$$a \approx E_0 E_s \sqrt{\frac{\rho}{\gamma}} R^{2.5}$$

where $E_0 E_s$ is the intensity of the biharmonic pump, $\rho$ is the drop density, $\gamma$ is the surface tension coefficient.

For the oscillation amplitude value of the order of the $10^{-2}$ R the pump intensity must be about $10^8$ W/cm$^2$. Due to the low absorption of the ruby laser radiation (the absorption coefficient of the virus suspension is less than $10^{-3}$ cm$^{-1}$, approximately like for water), the direct thermal effect of laser radiation on the system is absent. The influence of the ponderomotive interaction on the system under study with no absorption will be decisive. Taking into account the resonant nature of the interaction and, as a consequence, the relatively large amplitudes of the viral oscillations, biharmonic pumping can be regarded as an effective tool of selective influence on various biological systems.

## 4. Conclusion

In summary, we have experimentally registered SLFRS in CaMV suspension in Tris-HCl pH7.5 buffer with a high conversion efficiency. We also defined the energy and spectral characteristics of this process. We showed that the spectral characteristics of SLFRS can be used for a definition of the elastic properties of the systems under consideration. This is important for the investigation of biological systems. According to the fact that SLFRS can be excited in a large variety of nanoscaled systems and the frequency shift can be easily varied from several gigahertz till 1 terahertz SLFRS can be used as the source of biharmonic pumping for powerful selective and resonance impact on different biological systems.

**References**

[1] Szu-Chi Yang, Huan-Chun Lin, Tzu-Ming Liu et al 2015 *Sci. Rep.* **5** 18030
[2] Michels B, Dormoy Y, Cerf R, and Schulz J A 1985 *J. Mol. Biol.* **181** 103
[3] Cerf R, Michels B, Schulz J A et al 1979 *Proc. Natl.Acad. Sci. USA* **76** 1780
[4] Yan Y X, Gambel E B Jr and Nelson K A 1985 *J. Chem. Phys.* **83** 5391
[5] Tsen K-T, Tsen S W D, Chang C L et al 2007 *Virol. J.* **4** 50
[6] Tsen K-T, Tsen S W D, Chang C L et al 2007 *J. Phys.: Condens. Matter* **19** 322102
[7] Tsen K-T, Tsen S W D, Sankey O F and Kiang 2007 J G *J. Phys.: Condens. Matter* **19** 472201-1
[8] Karpova O V, Kudryavtseva A D, Lednev V N et al 2017 *Laser Phys. Lett.* **13** 085701.
[9] Ford L H 2003 *Phys.Rev. E* **67** 051924-1
[10] Murray D B and Saviot L 2005 *Physica E* **26** 417
[11] Balandin A and Fonoberov V 2005 *J. Biomed. Nanotechnol.* **1** 90
[12] Duval E, Boukenter A and Champagnon B 1986 *Phys. Rev. Lett.* **56** 2052
[13] Ivanda M, Babocsi K, Dem C et al 2003 *Phys. Rev. B* **67** 235329
[14] Montagna M 2008 *Phys. Rev. B* **77** 045418
[15] Tsen K T, Dykeman E C, Sankey O F et al 2006 *Virol. J.* **3** 79
[16] Dykeman E C, Sankey O F and Tsen K-T 2007 *Phys. Rev. E* **76** 011906
[17] Murray D B and Saviot L 2007 *J. Phys. Conf. Ser.* **92** 012036
[18] Tcherniega N V, Samoylovich M I, Kudryavtseva A D et al *2010 Opt. Lett.* **35** 300
[19] Tcherniega N V, Zemskov K I, Savranskii V V et al 2013 *Opt. Lett.* **38** 824
[20] Averyushkin A S, Baranov A N, Bulychev N A et al 2017 *Optics Communications* **389** 75
[21] Nikitin N, Trifonova E, Karpova O and Atabekov J 2013 *Microsc. Microanal.* **19** 808
[22] Petrova E, Nikitin N, Trifonova E et al 2015 *Biochimie* **115** 116
[23] Bykovskiĭ Yu A, Manykin E A, Nakhutin I E et al 1976 *Sov. J. Quantum Electron* **6 (1)** 84